\documentclass [a4paper,12pt] {article}
\usepackage{setspace, multicol, amsfonts, stmaryrd, txfonts, pifont, color, graphicx,pstricks,amssymb}

\newcommand {\eqref} [1] {(\ref {#1})}
\newcommand {\slsh} [1] {\not{\hbox{\kern-2pt${#1}$}}}

\def\drawbox#1#2{\hrule height#2pt
         \hbox{\vrule width#2pt height#1pt \kern#1pt
               \vrule width#2pt}
               \hrule height#2pt}

\def\Asym#1#2{\vcenter{\vbox{\drawbox{#1}{#2}
               \kern-#2pt       
               \drawbox{#1}{#2}}}}

\newcommand{\ap}{\ensuremath{\alpha^{\prime}}}
\newcommand{\app}{\ensuremath{\alpha^{\prime 2}}}
\newcommand{\nf}{\ensuremath{\mathcal{F}}}
\newcommand{\innit}{\!\!\int\!\!}
\newcommand{\ads}{\ensuremath{AdS_{\!5}}}
\newcommand{\adsxs}{\ensuremath{AdS_{\!5}\times S^5}}
\newcommand{\ylam}{\ensuremath{y_{\Lambda}}}
\newcommand{\wa}{\ensuremath{\mathcal{W}_1}}
\newcommand{\wb}{\ensuremath{\mathcal{W}_2}}
\newcommand{\none}{\ensuremath{\mathcal{N}\!\!=\!\!1}}
\newcommand{\nunez}{N$\mathrm{\acute{u}}\tilde{\mathrm{n}}$ez }

\newcommand {\beq} {\begin{equation}}
\newcommand {\eeq} {\end{equation}}
  \newcommand {\ber}{\begin{eqnarray*}}
  \newcommand {\eer} {\end{eqnarray*}}
\newcommand {\bea}{\begin{eqnarray}}
  \newcommand {\eea} {\end{eqnarray}}

\newcommand{\Dslash}{\,{\raise.15ex\hbox{/}\mkern-12mu D}}

\begin{document}


\begin{titlepage}

\begin{flushright}
\end{flushright}

\begin{center}
\vspace{1in}
\LARGE{Quantum Broadening of $k$-Strings\\from the AdS/CFT Correspondence}\\
\vspace{0.4in}
\large{Adi Armoni and Jefferson M. Ridgway}\\
\small{\texttt{a.armoni@swan.ac.uk},\, \texttt{pyjr@swan.ac.uk}}\\
\vspace{0.2in}
\large{\emph{Department of Physics, Swansea University}\\ 
\emph{Singleton Park, Swansea, SA2 8PP, UK.}\\}
\vspace{0.5in}
\end{center}

\abstract{We study the quantum broadening (width) of $k$-strings. We generalise
an old result by L\"uscher, M\"unster and Weisz to the case of a $k$-string, by using the gauge/gravity correspondence. When the fundamental QCD-string is replaced by a bound state of $k$ strings, the bound state is better described by a wrapped D-brane. We calculate the width of the $k$-string (the wrapped D-brane) in several AdS/CFT backgrounds by using a D-brane probe and find universally that  $\omega_k ^2~=~ {1\over \ 2\pi \sigma _k} \log {R / r}$.}
\end{titlepage}


\section{Introduction}
\label{introduction}

\paragraph{}In the confining phase of (Super-)Yang-Mills theories, the center of the gauge group is an unbroken global symmetry. One can classify the colour-singlet objects according to their charge under the center. In an SU($N$) gauge theory the center is $Z_N$. The possible charges under the center are $\exp i\, 2\pi k/N$ ($k$ is often called the $N$-ality).

Of particular interest are $k$-strings. They are the confining strings stretched between a quark -- anti-quark pair in a representation $R$ of $N$-ality $k$. It is possible to think about the $k$-string as a bound state of $k$ individual strings.
$k$-strings, being interesting non-perturbative objects in confining theories, are a subject of an intensive study in recent years. Most of the literature is dedicated to the calculation of their tension in various approaches: SUSY gauge theories \cite{Douglas:1995nw,Armoni:2003ji}, MQCD \cite{Hanany:1997hr}, the AdS/CFT correspondence \cite{Herzog:2001fq,Drukker:2005kx,Hartnoll:2006hr,Ridgway:2007vh} and lattice simulations \cite{DelDebbio:2001kz,Lucini:2001nv,DelDebbio:2001sj,Lucini:2004my}. The dependence on the quark representation was also studied recently \cite{Armoni:2003nz,DelDebbio:2003tk,Armoni:2006ri,Bringoltz:2008nd}. See \cite{Shifman:2005eb} for a short review.  

Another interesting question is what is the width of the $k$-string. The answer to this question, for the case of a fundamental QCD-string was provided a long time ago in the seminal paper of L\"uscher, M\"unster and Weisz \cite{Luscher:1980iy}. In their paper entitled {\em ``How Thick Are Chromoelectric Flux Tubes''}, the authors proposed to measure the thickness of the QCD-string, represented as a Wilson loop, by another small Wilson loop located at an orthogonal distance from the `original' Wilson loop. The role of the probe loop is to measure the chromoelectric field of the QCD-string at any given distance, $L$. Evaluating over all possible distances, L\"uscher et.al. found that the width is
\beq
 \omega ^2= {1 \over 2\pi \sigma } \log {R \over r} \, , \label{LMW}
\eeq
where $R$ is the quark  -- anti-quark separation, $r$, the radius of the probe loop, is a UV cut-off and $\sigma$ is the string tension. Surprisingly the string width depends on its size $R$. The QCD-string width was analysed by using the AdS/CFT correspondence in \cite{Greensite:2000cs}.

A first attempt at calculating the $k$-string width was made recently in \cite{Giudice:2006hw}. The conclusion of the authors is that the string width is $k$ independent. The conclusion was based on an analysis where the $k$-string and the probe were described by the Nambu-Goto action. In such a setup 
 the probe Wilson loop can interact with one individual constituent of the bound state at a time and therefore the width cannot depend on $k$.

In this note we calculate the width of the $k$-string by using the AdS/CFT correspondence. In this approach, in the limit $k/N={\rm fixed}$ and $N\rightarrow \infty$, the $k$-string becomes a wrapped D-brane. This is a universal property of various setups. It is either a wrapped D5-brane in a $AdS$ hard-wall model, or a wrapped D3-brane in both the Maldacena-\nunez and Klebanov-Strassler backgrounds. In our approach the probe Wilson loop will `see' the $k$-string as a single object, and will probe the chromoelectric field of this bound state and not of its individual constituents. Our result is
\beq
 \omega_k ^2= {1 \over  2\pi \sigma _k} \log {R \over r} \, ,
\eeq
namely that the generalisation of the formula \ref{LMW} is simply by replacing $\sigma$ with $\sigma _k$.

The paper is organised as follows: in section 2 we briefly review the calculation of \cite{Luscher:1980iy}. Section 3 is devoted to a discussion about how the result depends on the probe we use. In section 4 we present our calculation in the case of `an $AdS$ with a hard-wall' model. In section 5 we generalise our calculation to other confining backgrounds where the $k$-string is represented by a D3 brane.

\section{Fundamental string width review}
\label{f string width}
\paragraph{}The paper of L\"uscher et al. \cite{Luscher:1980iy} prescribed a method of determining the width of colour confinement flux tubes via the Nambu-Goto minimal surface of a Wilson loop correlator in flat space. It was proposed that this approach yielded the width of a flux tube associated with a fundamental string (e.g. a single quark -- anti-quark pair). The construction calls for two concentric circular Wilson loops, \wa \& \wb\, of unequal radii $R_1$ \& $R_2$, sitting in the plane of $x_1, x_2$, while being separated by a transverse distance $L$ in $x_3$. The minimal surface area is described by a minimised string world sheet that stretches between \wa \& \wb, namely a catenoid.
\paragraph{}To be able to determine the width of the string, the radius of \wb\, is taken to be very much smaller than that of \wa, effectively acting as a probe. \wb\, being taken to infinitesimal size can be thought of as a point operator, namely ${\rm tr} \, F_{\mu \nu}^2$, measuring the chromo electric field strength at a distance $L$ from the quark -- anti-quark pair. This probe is then evaluated over all possible distances from \wa.

\paragraph{}The connected Wilson loop correlator is related to the minimal surface in the following way

\begin{equation}
P(L) = \frac{\langle\wa \wb\rangle - \langle\wa\rangle\langle\wb\rangle}{\langle\wa\rangle}\propto e^{-\sigma A(L)}
\end{equation}

\paragraph{}The square of the string width subsequently follows from

\begin{equation}
\omega^2= \frac{\int_0^\infty P(L)L^2 dL}{\int_0^\infty P(L) dL}
\end{equation}

\paragraph{}In hard-wall $AdS$ (\adsxs with a cutoff in the radial $AdS$ direction) the fundamental correlator provides the same result as that of L\"uscher, with a suitably modified string tension, determined by the cutoff value, $\ylam$,

\begin{equation}
\omega^2_f = \ylam^2 \log[R_1/R_2]=\frac{1}{2 \pi\sigma_f}  \log[R_1/R_2]
\end{equation}

\paragraph{}To generalise this method to consider strings of higher representations, $k$-strings, it is proposed to replace the string world sheet by the world sheet of a D-brane wrapping a suitable manifold. The lowest energy $k$-string is in the anti-symmetric representation. In the case of string tension calculations, the antisymmetric $k$-string in $\adsxs$\, (and hard-wall $AdS$) is described by a D5-brane wrapping a 4-cycle inside the transverse $S^5$, while the remaining two directions act as a string in $AdS$ \cite{Hartnoll:2006hr}. The proposition is to replace the string in hard-wall $AdS$ with a D5-brane wrapped on the same 4-cycle as utilised in the string tension calculations \cite{Ridgway:2007vh}.
\section{Which probe to use ?}
\label{which probe}

\paragraph{}A preliminary question that must  be raised is what type of probe should be used in calculating the $k$-string width. The width will dependent upon the type of the probe, so the aim is to find the most `physical' probe.

 Consider the two point function 
\beq
 \langle \wa(k) \, \wb (f) \rangle _{\rm conn.} \, , \label{p1}
\eeq
namely a Wilson loop $\wa$ of $N$-ality $k$ and a small fundamental Wilson loop $\wb $. For simplicity assume that the representation of the Wilson loop $\wa$ is the $k$-th tensor product of the fundamental representation, namely that the above equation \eqref{p1} takes the form
 \beq
\langle \,\left[\wa(f) \right]^{\,k} \, \wb (f) \rangle _{\rm conn.} \, . \label{p2}
\eeq
At large-$N$, due to factorisation, \eqref{p2} becomes

 \beq
k \langle  \wa (f) \rangle ^{k-1} \langle \wa (f) \, \wb(f) \rangle _{\rm conn.} \, . \label{p3}
\eeq
Thus, if the probe Wilson loop is in the fundamental representation, it will necessarily `see' only one constituent of $\wa$ at a time. This leads to the conclusion that the width of $\wb$ is the same as the width of a Wilson loop in the fundamental representation \cite{Giudice:2006hw}.

Alternatively, consider a probe of $N$-ality $k$ (more precisely a $k$-th tensor product of the fundamental). In such a case the two point function \eqref{p1} is replaced by
\beq
 \langle \wa(k) \, \wb (k) \rangle _{\rm conn.} \, , \label{p4}
\eeq
which at large-$N$ takes the form of 
 \beq
 \langle\, \left[\wa(f)\right]^{\,k} \, \left[\wb (f)\right]^{\,k} \rangle _{\rm conn.} =  k! \langle \wa(f) \, \wb (f) \rangle _{\rm conn.} ^k \, . \label{p5}
\eeq
Here, the probe interacts simultaneously with all the constituents of the $k$-string and the resultant measurement is that the $k$-string is a factor of $k$ narrower than the fundamental string. Namely, $\sigma$ in \eqref{LMW} is replaced by $k \sigma$.

The above analysis is formulated in the limit where $k$ is fixed and $N$ is large. In such a limit the $k$ string is simply a collection of free $k$ strings. In this paper, a different limit  is focused upon, namely one where both $k$ and $N$ are large while $k/N$ is kept fixed. Motivated by the above analysis, a setup is considered where the probe is in the same representation as the `original' Wilson loop. This enables the width of the bound state to be measured, and not the width of one of its constituents.

\section{$k$-string width}
\label{k string width}

\paragraph{}In analogy with the fundamental string case, consider a D5 brane wrapping an $S^4 \!\subset\! S^5$, with the remaining two directions along $\tau$ \& $\sigma$, the string co-ordinates. The action of the brane is described by DBI \& Wess-Zumino parts;

\begin{equation}
S_{\!Bulk}=T_{\!D5} \innit d^6 \xi \sqrt{det(\mathcal{G} +\mathcal{F})} - i T_{D5} \innit d^6 \xi\, C_4 \wedge \mathcal{F}
\end{equation}

\paragraph{}The integration is performed over the $S^4 \!\subset\! S^5$, $\tau$ \& $\sigma$, with $T_{\!D5}$ as the brane tension, $ \mathcal{G}$ as the induced metric on the D5, $C_4$ the Ramond-Ramond 4-form potential that exists in the $S^4$\,, which satisfies $G_5=dC_4$. $\mathcal{F}=2\pi\ap F_{\tau \sigma}$, where $F_{\tau \sigma}$ is the quantised chromoelectric field strength that sits on the brane in $\tau,\sigma$ space.

\paragraph{}In addition to the bulk action, there is an additional momentum term due to the effects of the field strength at the boundary. The addition of this term provides a total action for the brane;

\begin{equation}
S_{\!Total}=S_{\!Bulk}-\mathcal{F} \frac{\partial}{\partial \mathcal{F}} S_{\!Bulk}
\end{equation}

\paragraph{}It is this total action that will describe the minimal area of the catenoid between \wa and \wb. This shall be employed to provide the `width' of the $k$-string.

\paragraph{}The metric under consideration is hard-wall $AdS$, so the calculation will be restricted to a flat space slice, $\varmathbb{R}^4$, in \ads\, at the cut off \ylam.  The $\varmathbb{R}^4\!\times\! S^5$ metric in Euclidean Poincar\'e co-ordinates is given as;

\begin{equation}
ds^2 = \frac{1}{\ylam^2} \left( dr^2 +r^2 d\phi^2 + dz^2 +z^2 d\chi^2\right) + d\theta^2 +\sin^2 \theta d \Omega_4^2
\end{equation}

\paragraph{}The orientation of the two loops is the same as that of L\"uscher, the fundamental case. The loops \wa\, \& \wb \, lie in $r,\phi$ space, separated in the $z$ direction with \wa\, at $z=0$, and \wb\, at $z=L$. Note that due to the angular symmetry of the problem, the distance between the centres of the Wilson loops does not depend on $\chi$. As the loops are concentric, the centre of each loops lies at $r=0$. At \wa\, $r=R_1$, and as $z$ increases, $r$ will interpolate towards $r=R_2$ at \wb, likely reaching some minimum in-between. $d \Omega_4^2$ represents the $S^4 \subset S^5$ which is wrapped by the brane, while the angle $\theta$ is the constant angle which the $S^4$ sits in the $S^5$, and is related to the string charge $k$.
\paragraph{}The Ramond-Ramond 4-form potential, $C_4$, in this co-ordinate system is of the form

\begin{equation}
C_4=\left(\frac 32 \theta -  \sin 2 \theta +\frac 1{8}\sin 4 \theta \right) d \Omega_4^2
\end{equation}
\paragraph{}Due to the symmetry of the system, let $\phi$ to be identified with $\sigma$, and allowed to vary across $[0, 2\pi]$. As the radius, $r$, of the catenoid varies with $z$, allow both $r$ and $z$ to be general functions of $\tau$.

\begin{equation}
r\rightarrow r(\tau), \,\,\,\,\,\,\,z\rightarrow z(\tau)
\end{equation}

\paragraph{}Using this string embedding, the bulk action is expressed as

\begin{equation}
S_{Bulk} = T_{\!D5} \innit d^6 \xi \left[ \sin^4 \theta \sqrt{\frac{r^2}{\ylam^4} (z^{\prime 2}+r^{\prime 2}) - 4 \pi \app F^2} +2 \pi \ap F G(\theta)\right]
\end{equation}

\paragraph{}With $T_{\!D5}$ as the D-brane tension, and $d^6 \xi = d\tau d\sigma d \Omega_4$. For simplicity $G(\theta)=\left(\frac 32 \theta -  \sin 2 \theta +\frac 1{8}\sin 4 \theta \right)$. The primes denote derivatives with respect to $\tau$. On a technical note, due to the euclidean signature, \nf\, is re-expressed in terms of $F$, where $\nf=2\pi \ap F_{\tau \sigma}=i\,2\pi \ap F$.

\paragraph{}In \adsxs\, it is known that $\theta=\theta_k=\mathrm{constant}$ is related to the string charge $k = \frac{\partial S_{\!Bulk}}{\partial F}$, by the following relation

\begin{equation}
k=\frac N \pi \left(\theta_k - \frac 12 \sin 2 \theta_k \right)
\end{equation}

\paragraph{}This relation can be shown to hold in this system also. $k = \frac{\partial S_{\!Bulk}}{\partial F}$ leads to the expression of $F$ in terms of $\theta_k$

\begin{equation}
2\pi \ap F= \cos \theta_k \frac r {\ylam^2} \sqrt{z^{\prime 2}+r^{\prime 2}}
\end{equation}

\paragraph{}As mentioned earlier, there exists a momentum term due to the electric field strength $F$ which must be added to the bulk action to provide the total action of the system (Addition of this term is equivalent to taking the Hamiltonian of the bulk Lagrangian).

\begin{equation}
S_{\!Total} = S_{Bulk}- k F
\end{equation}

\paragraph{}Applying the expression for $F$ \& $k$ into the total action, using $T_{D5}=\frac{N}{8 \pi^4 \ap}$,  and integrating over the $S^4$, the action simplifies to

\begin{equation}
S_{Total}= \frac{2N}{3\pi \ap}\frac 1{2 \pi \ap} \sin^3 \theta_k \innit d\tau d\phi\, \frac r {\ylam^2} \sqrt{z^{\prime 2}+r^{\prime 2}}
\end{equation} 

\paragraph{}The classical equations of motion for $r$ \& $z$ are

\begin{eqnarray}
\sqrt{z^{\prime 2}+r^{\prime 2}}-\left(\frac{r \, r^{\prime}}{\sqrt{z^{\prime 2}+r^{\prime 2}}}\right)^{\prime}&=&0\\
\frac{r \, z^{\prime}}{\sqrt{z^{\prime 2}+r^{\prime 2}}}&=&m
\end{eqnarray}

\paragraph{}$m$ is defined as a constant in $\tau$. It is now appropriate to make a gauge choice: As both $\tau$ and $z$ increase monotonically, let $z(\tau)=\tau=z$. Using this choice, the equations of motion combine and simplify, and using  the fact $m$ is now constant in $z$;

\begin{equation}
1+ r^{\prime 2} -r r^{\prime \prime}=0
\end{equation}

This is solved by 

\begin{equation}
r(z)=B \cosh \left[\frac{z-z_0}B\right]
\end{equation}

where $z_0$ is defined as the value of $z$ at the minimum radius of the catenoid, and $B$ is a constant. Applying this solution to the total action, and integrating over $z\in[0,L]$, \& $\phi\in[0,2\pi]$.

\begin{eqnarray}
S_{Total}&=& \frac{2N}{3\pi} \frac 1{2 \pi \ap} \frac{B}{\ylam^2}\sin^3 \theta_k \innit dz d\phi \cosh^2\left[\frac{z-z_0}{B}\right]\\
&=& \frac{N}{3\pi\ap} \frac{B}{\ylam^2}\sin^3 \theta_k\left( L- \frac B2\left[\sinh\left(\frac{2(z_0-L)}B\right)-\sinh\left(\frac{2z_0}B\right)\right]\right)
\end{eqnarray}

\paragraph{}From the solution for $r(z)$, the expression for $z_0$ can be found by using the boundary conditions for $z$ \& $r$.

\begin{eqnarray}
R_1= B \cosh\left[\frac{z_0}B\right],\,\,\,\,R_2 = B \cosh \left[\frac{L-z_0}B\right]\\
z_0=\frac 12 \left[L-B\left(\mathrm{arccosh}\left[\frac{R_1}{B}\right]+\mathrm{arccosh}\left[\frac{R_2}{B}\right]\right)\right]
\end{eqnarray}

\paragraph{}An expression for $B$ cannot be determined in an analytic fashion, therefore an approximation is required.

Stepping back for a moment to consider the model, for the string width \wb\, must be considered as a probe loop, and thus must be very small, as outlined earlier. Therefore consider the limit of $R_1$ \& the ratio $R_1/R_2$ becoming large. In such a limit, the value of $B$ is approximated to leading order as

\begin{equation}
B= \frac{L}{\log [R_1/R_2]} +O\,(\log [R_1/R_2])^{-2}
\end{equation}

\paragraph{}Substituting into the total action the expressions for $B$ and $z_0$

\begin{equation}
S_{\!Total}=\frac{N}{3\pi\ap} \frac{1}{\ylam^2}\sin^3 \theta_k\left[(R_2^2-R_1^2) + \frac{L^2}{\log[R_1/R_2]}\right]
\end{equation}

\paragraph{}As only the second term has a dependence on the loop separation, $L$, when the string width is computed, in the numerator the exponent will cause the first term to cancel with an identical term from the denominator, thus only the second term is relevant. The width is calculated as

\begin{eqnarray}
\omega^2_k &=& \frac{\int^\infty_0 e^{-S_{\!Total}}L^2 dL}{\int^\infty_0 e^{-S_{\!Total}} dL}\\
&=&\frac{3\pi\ylam^2\ap}{2N\sin^3 \theta_k} \log[R_1/R_2]\\
&=&\frac{1}{2\pi \sigma_k} \log[R_1/R_2]
\end{eqnarray}

\paragraph{}This is precisely the result of the fundamental string with $\sigma_f\rightarrow\sigma_k$. For a general $k$ \& $k^{\prime}$, the ratio would give

\begin{equation}
\frac{\omega^2_k}{\omega^2_{k^{\prime}}}=\frac{\sin^3 \theta_{k^{\prime}}}{\sin^3 \theta_k}
\end{equation}

\section{$k$-string widths in $\none$ SYM gravity duals}

\paragraph{}As interesting as it is to look at an $AdS$ hard-wall background, it would be more meaningful to study the flux tube width in a theory with a resemblance to QCD (albeit with only one flavour), namely \none\, super Yang-Mills. The gravity dual of \none\, SYM can be approached from a number of directions: non-critical type IIA string theory; a D5 geometry back-reaction construction of Chamseddine-Volkov \cite{Chamseddine:1997mc}, later interpreted by Maldacena-\nunez \cite{Maldacena:2000yy}  as the dual of \none\,, and the deformed conifold system of Klebanov-Strassler, in IIB \cite{Klebanov:2000hb}. Focus will be upon the IR limit of the Maldacena-\nunez (MN) and Klebanov-Strassler (KS) models, as in this IR limit, both backgrounds become identical, up to some multiplicative constants.

\paragraph{}The string tensions for antisymmetric $k$-strings were calculated for both the MN and KS systems in the far IR \cite{Herzog:2001fq}. This was conducted by wrapping a D3 brane on $\sigma$ \& $\tau$ in the 4d spacetime, with the remaining two directions wrapping a 2-cycle in the transverse space. In both MN and KS, a 3-cycle can be chosen with the form of an $S^3$, and it is then simple to wrap the D3 on the $S^2\!\subset S^3$. In the NS-NS sector, the angle $\Psi$ at which the $S^2$ sits in the $S^3$, is dependent of the string charge $k$ which arises from the electric field, $F_2$ that lives on the brane in the $S^2$. Also in the NS-NS sector there exists the anti-symmetric tensor $B_2$. It was later shown that the same string tension can be determined in the S-dual, R-R sector, with the electric field in $\sigma$ \& $\tau$, and a $C_2$ confined to the $S^2$ \cite{Ridgway:2007vh}. 

\paragraph{}In this section it will be shown that the $k$-string width can be calculated in the MN background (with R-R flux) in the IR limit using the method presented in the previous section\footnote{Although the R-R flux background has a greater cross over to the string width calculation of hard wall $AdS$, the width can easily be calculated in the background with NS-NS flux also via S-duality.}, but replacing the D5 brane wrapping an $S^4\subset S^5$, with a D3 brane wrapping an $S^2$ in a 3-cycle of the transverse space.

The 10 dimensional cylindrical Euclidean MN metric can be expressed as follows

\begin{equation}
ds^2_{10}=(dr^2+r^2 d\eta^2 +dz^2 +dx_4^2) +    N \alpha^{\prime} \left(dy^2 +e^{2h(y)} \left(d\theta_1^2 + \sin \theta_1 d \phi_1^2\right) + \frac{1}{4}\,(\omega_i - A^i)^2 \right)
\end{equation}

Where: 
\begin{equation}
A^1=-a(y)d\theta_1,\;\;\;\;\; A^2=a(y)\sin \theta_1 d\phi_1, \;\;\;\;\; A^3=-\cos \theta_1 d\phi_1.
\end{equation}
\begin{eqnarray}
\omega_1&=&\cos \psi d\theta_2 +\sin\psi \sin\theta_2 d\phi_2,\\\nonumber
\omega_2&=&-\sin\psi d \theta_2 +\cos \psi \sin \theta_2 d\phi_2,\\\nonumber
\omega_3&=&d\psi +\cos \theta_2 d\phi_2
\end{eqnarray}

with $a(y)$ and $h(y)$  functions dependent on the `radial' co-ordinate $y$.  The topology of the transverse space is of two 2-spheres, $S^2_1$ \& $S^2_2$, with an $S^1$ fibration between them. The angles $\theta_1$, $\phi_1$  \& $\theta_2$, $\phi_2$  parametrise the  $S^2_1$ \& $S^2_2$ respectively, while the fibered $S^1$ by $\psi$. Along with the metric there exists a $C_2$ Ramond-Ramond potential, which obeys $F_3 = d C_2$, and is given by:

\begin{eqnarray}
C_2& =& \frac{\textstyle N \ap}{\textstyle 4}[\psi\,(\sin \theta_1 d\theta_1 \wedge d\phi_1-\sin \theta_2 d\theta_2 \wedge d\phi_2)\\\nonumber &&\,\,\,\,\,\,\,\,\,\,- \,\cos \theta_1\cos \theta_2 d\phi_1 \wedge d\phi_2 - (d\theta_1 \wedge \omega_1 - \sin \theta_1 d \phi_1 \wedge \omega_2 )]
\end{eqnarray}
 
The IR limit is defined as $y\rightarrow 0$, causing the functions $a(y)\rightarrow 1$ and $e^{2h(y)} \rightarrow 0$. Making the choice $\theta \equiv \theta_1=\theta_2$, $\phi \equiv\phi_1 = 2\pi- \phi_2$, \& $\psi \rightarrow 2\Psi +\pi$, a 3-cycle, $S^3$ is selected from the transverse space, causing the metric and  $C_2$ to reduce to;

\begin{equation}
ds^2 = (dr^2+r^2 d\eta^2 +dz^2 +dx_4^2) + N \alpha^{\prime}\left[ dy^2 + d\Psi^2 + \sin^2\Psi (d\theta^2 +\sin^2 \theta d\phi^2) \right]\end{equation}
\begin{equation}
C_2= N \alpha^{\prime}\left( \Psi - \frac 12 \sin 2\Psi\right)\sin \theta d\theta \wedge d\phi 
\end{equation}

\paragraph{}The action is that of the previous section, namely the DBI, Wess-Zumino and chromoelectric field strength momentum terms. With \wa\, and \wb\, set in $r$, $\eta$ space, while being separated in $z$, the D3 is wrapped along $\sigma$ \& $ \tau$, with $\eta \rightarrow \sigma$, $r\rightarrow r(\tau)$ and $z\rightarrow z(\tau)$, while the remaining two directions are wrapped upon an $S^2$ in the transverse 3-cycle, $S^3$. Turning on an electric field strength $\mathcal{F}=2\pi \ap F_{\tau \sigma}=2 \pi \ap i F$, and allowing the string charge $k=\frac{\partial S_{\!Bulk}}{\partial F} = \frac{\Psi N}\pi $, the total action becomes

\begin{eqnarray}
S_{\!Total}&=&S_{\!DBI} + S_{\!WZ} - \mathcal{F} \frac{\partial }{\partial \mathcal{F}}\left(S_{\!DBI} + S_{\!WZ}\right) \\\nonumber
&=& T_{\!D3} \!\!\int\!\! d^4 \xi \left[\sqrt{\mathrm{det} \,(\cal{G} +\mathcal{F})} - i C_2 \wedge \mathcal{F}\right] - \mathcal{F} \frac{\partial }{\partial \mathcal{F}}\left(S_{\!DBI} + S_{\!WZ}\right)\\
&=&\frac{N}{2\pi^2 \ap}\,\sin\frac{\pi\, k}{N}\!\!\int\!\!d\eta\, d\tau\, r\sqrt{z^{\prime 2}+r^{\prime 2}}
\end{eqnarray}

Again, primes denote derivatives with respect to $\tau$, and $T_{\!D3}=1/(2\pi)^3 \app$. The equations of motion for $r$ and $z$ are identical to those from the hard-wall case, implying that the solution and gauge choice for $r$ and $z$ respectively will directly apply to this computation. The systematics of the $k$ dependence in the transverse space, and those minimal area in 4d spacetime, factorise completely and seem not to influence each other. 

\paragraph{}Using the solution for $r$, the total action simplifies to become
\begin{equation}
S_{\!Total}=\frac{B N}{2\pi^2 \ap}\,\sin\frac{\pi\, k}{N}\!\!\int\!\!d\eta\, dz\,  \cosh^2 \left[\frac{z-z_0}{B}\right]
\end{equation}

\paragraph{}Integrating  over $z$ \& $\eta$ ($z \in [0,L]$ \& $\eta \in [0,2\pi]$), using the boundary conditions for $r$ \& $z$ to eliminate $z_0$, and taking the limit where $R_1$, \& the ratio $R_1/R_2$, both become very large, the total action to leading order is expressed as

\begin{equation}
S_{\!Total}=\frac{ N}{2\pi \ap}\,\sin\frac{\pi\, k}{N}\left[(R_2^2-R_1^2) + \frac{L^2}{\log[R_1/R_2]}\right]
\end{equation}

\paragraph{}Calculation of the string width provides a result of the same form as for the hard-wall $AdS$, and ultimately the fundamental string case.

\begin{eqnarray}
\omega^2_k &=&\frac{\pi\,\ap}{N\sin\frac{\pi\, k}{N}} \log[R_1/R_2]\\
&=&\frac{1}{2\pi \sigma_k} \log[R_1/R_2]
\end{eqnarray}

\paragraph{}The equivalent calculation in the KS background provides a string width of the exact same form; namely $\propto 1/\sigma_k$. It seems from these calculations, and a generalised calculation, that the string widths in a confining background will be of the general form of the reciprocal of the string tension. For a general $k$ \& $k^{\prime}$, the ratio becomes

\begin{equation}
\frac{\omega^2_k}{\omega^2_{k^{\prime}}}=\frac{\sin\frac{\pi\, k^{\prime}}{N}}{\sin\frac{\pi\, k}{N}} \, .
\end{equation}

\section*{Acknowledgements}

We wish to thank D. Berenstein, N. Drukker, C. \nunez, A. Rago, M. \"Unsal and K. Zarembo for fruitful discussions. Special thanks to F. Gliozzi for comments on a draft version of the paper. A.A. is supported by the PPARC advanced fellowship award. J.R. is sponsored by a PPARC studentship.


\end{document}